\documentclass[epj]{svjour}
 \usepackage{lineno} 
 \usepackage{color}
 \definecolor{MPPgreen}{RGB}{0,128,112}
\usepackage{graphicx}
\usepackage{dcolumn}
\usepackage{bm}
\usepackage{amssymb}
\usepackage{wasysym}

\begin{document}
\bibliographystyle{spphys} 

\title{Results on MeV-scale dark matter from a gram-scale cryogenic calorimeter operated above ground}


\author{
G.~Angloher\inst{1} \and
P.~Bauer\inst{1} \and
A.~Bento\inst{1,8} \and
C.~Bucci\inst{2} \and
L.~Canonica\inst{2,9} \and
X.~Defay\inst{3} \and
A.~Erb\inst{3,10} \and
F.~v.~Feilitzsch\inst{3} \and
N.~Ferreiro~Iachellini\inst{1} \and
P.~Gorla\inst{2} \and
A.~G\"utlein\inst{4,5} \and
D.~Hauff\inst{1} \and
J.~Jochum\inst{6} \and
M.~Kiefer\inst{1} \and
H.~Kluck\inst{4,5} \and
H.~Kraus\inst{7} \and
J.-C.~Lanfranchi\inst{3} \and
A.~Langenk\"amper\inst{3} \and
J.~Loebell\inst{6} \and
M.~Mancuso\inst{1} \and
E.~Mondragon\inst{3} \and
A.~M\"unster\inst{3} \and
L.~Oberauer\inst{3,}\thanks{Associated with the CRESST collaboration for this work.}  \and
C.~Pagliarone\inst{2} \and
F.~Petricca\inst{1} \and
W.~Potzel\inst{3} \and
F.~Pr\"obst\inst{1} \and
R.~Puig\inst{4,5} \and
F.~Reindl\inst{1,}\thanks{Current address: INFN - Sezione di Roma 1, I-00185 Roma, Italy} \and
J.~Rothe\inst{1} \and
K.~Sch\"affner\inst{2,11} \and
J.~Schieck\inst{4,5} \and
S.~Sch\"onert\inst{3} \and
W.~Seidel\inst{1,}\thanks{Deceased} \and
M.~Stahlberg\inst{4,5} \and
L.~Stodolsky\inst{1} \and
C.~Strandhagen\inst{6} \and
R.~Strauss\inst{1,}\thanks{Corresponding author: strauss@mpp.mpg.de} \and
A.~Tanzke\inst{1} \and
H.H.~Trinh~Thi\inst{3} \and
C.~T\"urko\v{g}lu\inst{4,5} \and
M.~Uffinger\inst{6} \and
A.~Ulrich\inst{3} \and
I.~Usherov\inst{6} \and
S.~Wawoczny\inst{3} \and
M.~Willers\inst{3} \and
M.~W\"ustrich\inst{1} \and
A.~Z\"oller\inst{3} \newline \newline (The CRESST Collaboration)
}

\institute{
Max-Planck-Institut f\"ur Physik, D-80805 M\"unchen, Germany \and
INFN, Laboratori Nazionali del Gran Sasso, I-67010 Assergi, Italy \and
Physik-Department and Excellence Cluster Universe, Technische Universit\"at M\"unchen, D-85747 Garching, Germany \and
Institut f\"ur Hochenergiephysik der \"Osterreichischen Akademie der Wissenschaften, A-1050 Wien, Austria \and
Atominstitut, Vienna University of Technology, A-1020 Wien, Austria \and
Eberhard-Karls-Universit\"at T\"ubingen, D-72076 T\"ubingen, Germany \and
Department of Physics, University of Oxford, Oxford OX1 3RH, United Kingdom \\ \and
Also at: Departamento de Fisica, Universidade de Coimbra, P3004 516 Coimbra, Portugal \and
Also at: Massachusetts Institute of Technology, Cambridge, MA 02139, USA \and
Also at: Walther-Mei\ss{}ner-Institut f\"ur Tieftemperaturforschung, D-85748 Garching, Germany \and
Also at: GSSI-Gran Sasso Science Institute, 67100, L'Aquila, Italy
}





\date{\today}

\abstract{
Models for light dark matter particles with masses below 1\,GeV/c$^2$ are a natural and well-motivated alternative to so-far unobserved weakly interacting massive particles. Gram-scale cryogenic calorimeters provide the required detector performance to detect these particles and extend the direct dark matter search program of CRESST. A  prototype 0.5\,g sapphire detector developed for the $\nu$-cleus experiment  has achieved an energy threshold  of $E_{th}=(19.7\pm 0.9)$\,eV. This is one order of magnitude lower than for previous devices and independent of the type of particle interaction.  The result presented here is obtained in a setup above ground without significant shielding against ambient and cosmogenic radiation. Although operated in a high-background environment, the detector probes a new range of light-mass dark matter particles previously not accessible by direct searches. We report the first limit on the spin-independent dark matter particle-nucleon cross  section for masses between  140\,MeV/c$^2$ and 500\,MeV/c$^2$. 
}

\titlerunning{Results on MeV-scale dark matter\dots}

\maketitle

\section{Introduction}
\label{intro}
The origin of dark matter (DM) is still unknown, despite plenty of compelling observational evidence \cite{Bertone2005279}. The most popular model for DM in recent years suggests the existence of weakly interacting massive particles (WIMPs) which are produced in the early Universe \cite{Jungman1996195}. Such particles are in particular appealing since the correct relic density of DM can be produced with a thermally-averaged annihilation cross section
$\langle \sigma v\rangle $ 
   which is amazingly close to the weak scale. This natural explanation is referred to as the ``WIMP miracle'' and has been the main driver for DM physics. Minimal supersymmetric (SUSY) models  provide an attractive candidate for WIMPs: the neutralino often assumed as the lightest supersymmetric particle with a mass of $\mathcal{O}$(1\,GeV/c$^2$) to $\mathcal{O}$(1\,TeV/c$^2$) \cite{Ellis2000181}. Lee and Weinberg constrain the mass of WIMPs to $m_\chi\geq2\,$GeV \cite{PhysRevLett.39.165}. Lighter particles would have a too small annihilation cross section  $\langle \sigma v\rangle \propto \frac{m_\chi^2}{m_Z^4}$ ($m_Z$: mass of the Z-boson)  which  would lead to a too large relic abundance of DM and, thus, an overclosure of the Universe.  Therefore, most experimental DM searches in the past decades focused on GeV-scale particles.

Despite enormous experimental effort, no clear and unambiguous signature for DM has been  found by direct, indirect or accelerator searches so far. 

Due to this lack of evidence, models which provide alternatives to WIMPs should be probed experimentally. In the past years, rising interest is drawn to light DM (\textit{l}DM) which extends the allowed mass range of  DM particles  to $\mathcal{O}$(keV). Asymmetric DM \cite{PhysRevLett.68.741,asymmReview,PhysRevD.79.115016}, scalar DM particles \cite{Boehm2004219,Boehm_Silk_ensslin} and hidden sector DM \cite{PhysRevLett.101.231301} are  examples for  promising and naturally motivated \textit{l}DM models. These theories are compatible with the observed  relic density (Lee-Weinberg argumentation), the constraints from existing DM searches, in particular indirect searches which are sensitive to \textit{l}DM, and limits from particle physics experiments.

While searches for axions probe particle masses of keV-scale and below, the mass range between 1\,MeV and 1000 MeV is only weakly constrained by \textit{l}DM-electron scattering \cite{PhysRevLett.109.021301} or via Bremsstrahlung emission in nuclear recoils \cite{PhysRevLett.118.031803,McCabe:2017rln}. 

We report here  results from the \textit{$\nu$-cleus} 0.5\,g prototype detector \cite{cnns_letter} which enables for the first time to directly probe  nuclear recoils induced by DM particles with masses below 500\,MeV/$c^2$.   

\section{A gram-scale cryogenic calorimeter}
\subsection{Detector technology}
Cryogenic  detectors are sensitive to the temperature rise induced by a particle interaction. CRESST-type detectors use single crystals (e.g. CaWO$_4$ or Al$_2$O$_3$) equipped with transition-edge-sensors (TES) made of thin tungsten films. The TES is usually directly coupled to the heat sink by a gold bond wire. Due to the operation at very low temperatures of $\sim10$\,mK and the resulting electron-phonon decoupling in the thermometer film, such a cryogenic calorimeter is mainly sensitive to non-thermal phonons \cite{Probst:1995fk}.  Very low energy thresholds, a high dynamic range (of up to 10$^5$) and a particle-type independent detector response are the main advantages of these devices. State-of-the-art  CaWO$_4$ detectors   of 300\,g operated in CRESST-II reach thresholds of $\sim300$\,eV \cite{Angloher:2015ewa}.  

The energy threshold of cryogenic detectors depends strongly on the mass $M$ of the crystal which can be quantified by a scaling law discussed in \cite{cnns_letter}. This  describes  nicely the performance of existing CaWO$_4$ and Al$_2$O$_3$ detectors and can be used to predict the sensitivity for  detectors of various sizes, materials and geometries \cite{cnns_long}.  In case of cubic crystals the energy threshold scales as $E_{th}\propto M^{2/3}$ suggesting that thresholds of 10\,eV and below are in reach for detectors with masses of $\mathcal{O}$(1\,g) \cite{cnns_letter}. 
\subsection{The prototype calorimeter}
The first prototype calorimeter made of Al$_2$O$_3$  has a size of 5$\times$5$\times$5\,mm$^3$ and a mass of $0.49$\,g. One side of the optically polished crystal is equipped with a specifically developed TES. 
\begin{figure}
\centering
\includegraphics[width=0.45\textwidth]{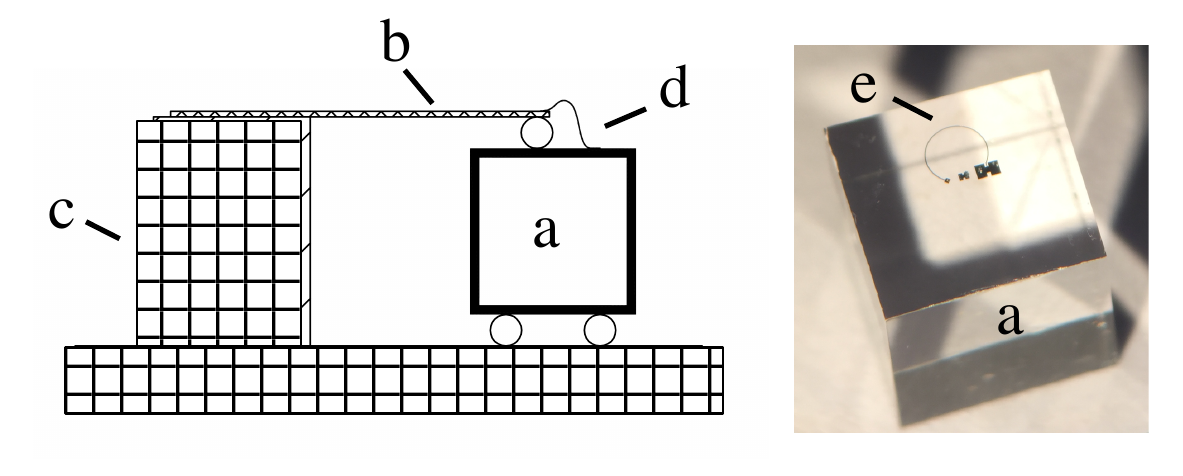}
\caption{Left: Schematic drawing of the detector crystal (a) and the copper holder (c). The crystal is placed on Al$_2$O$_3$ balls and is pressed from top by a flexible bronze clamp~(b). Bond wires~(d) provide electrical and thermal connection. Right: Picture of the prototype Al$_2$O$_3$ crystal (a) with the specially designed TES (e). For details see text.  }
\label{fig:detector}
\end{figure}
 The design is similar to the one used for CRESST light detectors, but adjusted to the size of the crystal \cite{cnns_long}. The sensor consists of a 200\,nm thick tungsten film with an area of 0.0061\,mm$^2$ and an aluminium phonon collector of thickness 1\,$\mu$m and an area of 0.15\,mm$^2$. It is weakly coupled to the heat bath via a gold stripe and a bond wire  (heat conductance: $\sim10$\,pW/K at 10\,mK). The dimension of the TES and the strength of the heat link are chosen such that the life time of non-thermal phonons in the detector $\tau_n$ is much shorter than the decay time of the pulses. The former depends on size, geometry and material of the crystal and the TES, the latter is determined by the strength of the heat link to the thermal bath.  A detector designed in this way operates in the calorimetric mode \cite{Probst:1995fk}, i.e. the phonons impinging on the film are integrated over the duration of the non-thermal phonon signal\footnote{Detectors operating in the bolometric mode, in contrast, measure the flux of phonons through the TES \cite{Probst:1995fk}.}.  A separate gold film sputtered on the crystal acts as an ohmic heater. A current applied to it heats the detector to the desired position in the superconducting phase transition. Artificial pulses of high energy (called control pulses \cite{Angloher2009270}) which saturate the detector response are injected into the heater (here: every 10\,s). Those allow to control and stabilize the operating point. In addition, artificial pulses of discrete, lower energies (called test pulses) are injected periodically  to continuously monitor the detector response  (here: every 60\,s).

A  fit of the pulse model \cite{Probst:1995fk} to the pulses of the prototype gram-scale detector confirms its calorimetric operation. The non-thermal phonon life-time in the crystal is $\tau_n=(0.30\pm0.01)$\,ms and the dominant decay time is found to be $\tau_{dec}=(3.64\pm0.01)$\,ms (for the non-thermal pulse component). 

Fig. \ref{fig:detector} (left) shows how the prototype detector is arranged in the experimental setup. 
The cubic crystal is installed on a copper plate and placed on three sapphire spheres (\diameter  1\,mm), pressed from the top with  another sphere attached to a flexible bronze clamp.  The sensor is placed on the top surface where it is electrically and thermally contacted to copper bond pads on the clamp via Al and Au wire bonds, respectively.  At a distance of about 2\,cm from the crystal, a $^{55}$Fe calibration source (activity $\sim$0.2\,Bq) is installed. 

\subsection{Experimental setup}
The prototype is operated in a cryostat of the Max-Planck-Institute for Physics in Munich in a surface building. A commercial dilution refrigerator from Oxford Instruments (Kelvinox400) is used with a standard helium dewar. The experimental volume is surrounded by a 1\,mm copper thermal screen, but no special care on radiopurity is taken in this setup. No dedicated shielding  against ambient radiation is installed within or outside the cryostat. The ceiling of the room is made of $\sim30\,$cm concrete, therefore a moderate reduction of the hadronic component of cosmogenic radiation is expected. In the present setup, no muon veto or other anti-coincidence detectors are installed.  

The TES is biased with a constant current (1.0\,$\mu$A) and read out with DC-SQUIDs (Jessy from Supracon).  The output of the SQUID electronics is fed to a low-noise voltage amplifier (SR560 from Stanford Research Electronics). 
For data-taking, two systems have been used in parallel: 1)~the standard CRESST data acquisition \linebreak(DAQ), see e.g. \cite{Angloher2009270}, with a hardware trigger unit and 16\,bit transient digitizer, and 2) a continuous data-taking system based on a 16\,bit digitizer from National Instruments (NI USB-6218 BNC). The latter was used to record a continuous stream of the entire measurement with a sampling frequency of 20\,kHz. Offline, a software trigger is applied to the data.



\section{Measurement}
The first measurement of the prototype gram-scale calori-\ meter in the MPI cryostat was primarily intended  to de-\ monstrate an energy threshold in the 10\,eV regime  for the  \textit{$\nu$-cleus} experiment \cite{cnns_long}. Since a threshold of $\sim20$\,eV was reached (see \cite{cnns_letter} and below), one order of magnitude lower than previous CRESST results \cite{Angloher:2015ewa},  this first data can be used to search for new physics:  a previously un-explored region of parameter space for DM-particle nucleus scattering can be explored (see section \ref{sec:results}). In this section we  review the energy calibration and threshold analysis which is presented in detail in \cite{cnns_letter}, and focus on dedicated aspects of the DM analysis.
\subsection{Energy calibration}
The measurement with the  0.49\,g Al$_2$O$_3$ prototype calori-\ meter had a total run-time of 5.31\,h corresponding to an exposure of 0.11\,g-days. The  $^{55}$Fe calibration source  was placed in the cryogenic setup during the entire run. The dominant K$_\alpha$ line ($E_{lit}=5.895\,$keV) of $^{55}$Mn is used to calibrate the energy spectrum.  The linearisation of the pulse response is established by a truncated template fit which uses the pulse shape infomation only in the linear region of the detector response \cite{cnns_letter}.  The method is well established and successfully used in previous CRESST ana-\ lyses \cite{Angloher:2014myn,Angloher:2015ewa}. The fit reproduces the   K$_\beta$ line of $^{55}$Mn  at an energy of $E_{obs}=(6.485\pm0.017)\,$keV), well in agreement with the literature value of 6.490\,keV \cite{cnns_letter}.  The robustness of the linearisation at lower energies  is studied and a moderate systematic error of 1.1\,\% on the energy calibration  is derived \cite{cnns_letter}.

\begin{figure}
\centering
\includegraphics[width=0.5\textwidth]{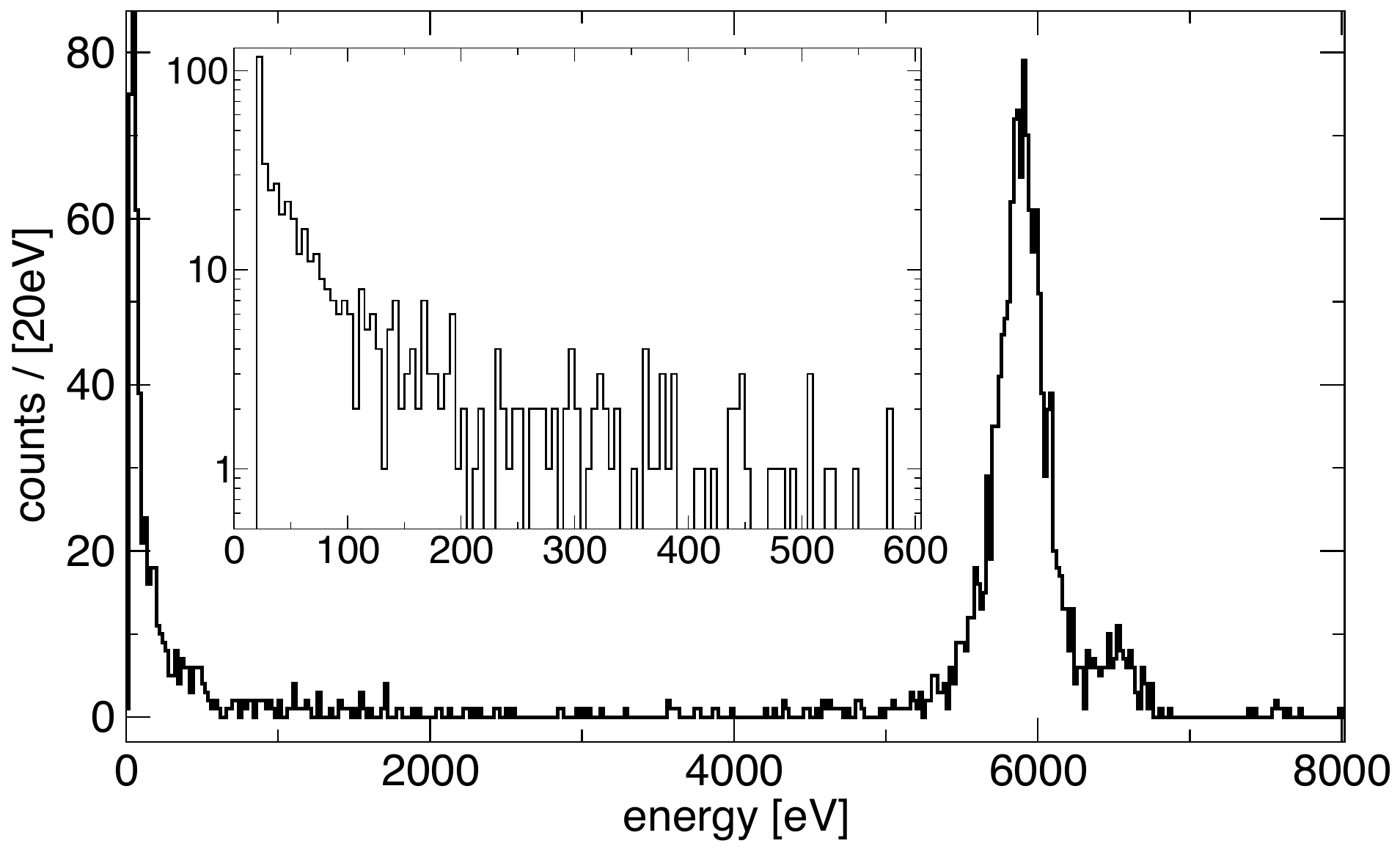}
\caption{Total energy spectrum of the 5.3\,h measurement in presence of the $^{55}$Fe x-ray source with peaks at 5.90 and 6.49\,keV. The inset shows the events in the region-of-interest for DM search from the energy threshold of 19.7\,eV  to 600\,eV (binning 5\,eV). No data quality cuts are applied. }
\label{fig:spectrum}
\end{figure}

Fig. \ref{fig:spectrum} (main frame) shows the final energy spectrum with the dominant x-ray lines of $^{55}$Mn. Above 7\,keV, a constant background rate of $\sim1.2\cdot10^{5}$\,counts/[kg keV day] is observed, while from $\sim1$\,keV towards lower energies the spectrum is significantly rising to a rate of about $10^8$\,counts/[kg keV day] (see discussion below). 

\subsection{Software trigger}

\begin{figure}
\centering
\includegraphics[width=0.45\textwidth]{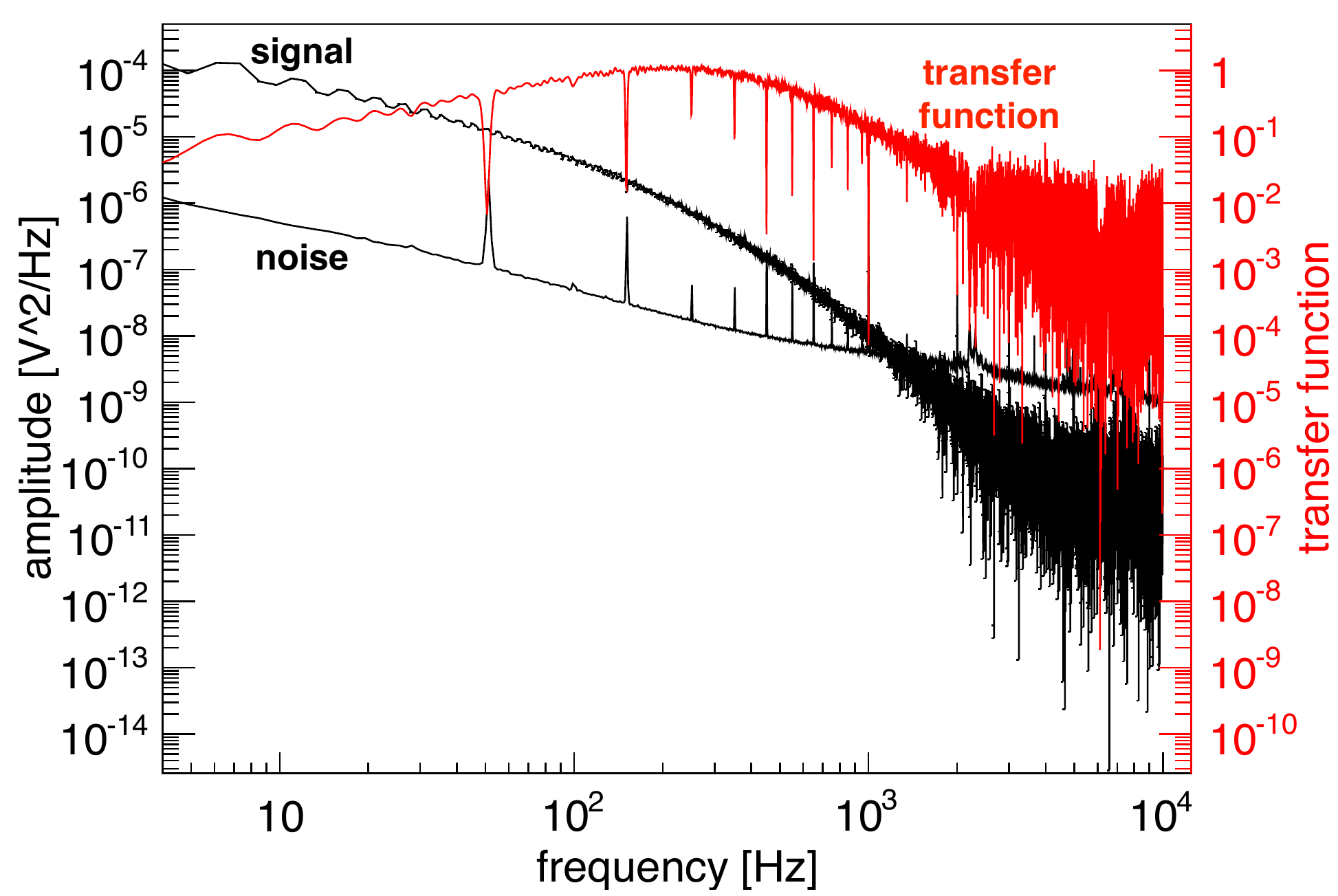}
\caption{Left axis: measured power spectra of randomly selected noise traces  and of the template  pulse (signal). Right axis: the corresponding filter transfer function. See text for details. }
\label{fig:optimum}
\end{figure}

For the evaluation of the pulse amplitudes in the linear region of the  detector response (up to 600\,eV), the optimum filter is used \cite{Gatti:1986cw,Piperno:2011fp}. This method improves the reconstruction of a known signal in the presence of noise with a measured power spectrum and typically gives significantly better results than the template fit.  The filter transfer function is calculated by the ratio of the power spectra of the template pulse and the noise. This function is used to weight the spectral components of the data sample  (in the frequency domain) according to the respective signal-to-noise-ratio. Fig. \ref{fig:optimum} shows the measured power spectra (left axis) and the resulting transfer function (right axis) of the optimum filter used for the analysis of the prototype calorimeter.  After being applied in the frequency space, the filter output is transformed back to time domain and normalized to match the maximum of the original pulse.  In case of this measurement, a baseline noise of $\sigma_b=(3.74\pm0.21)$\,eV \cite{cnns_letter} is reached with the optimum filter, a factor of 1.7 smaller compared to that obtained with the template fit. 

\begin{figure}
\centering
\includegraphics[width=0.45\textwidth]{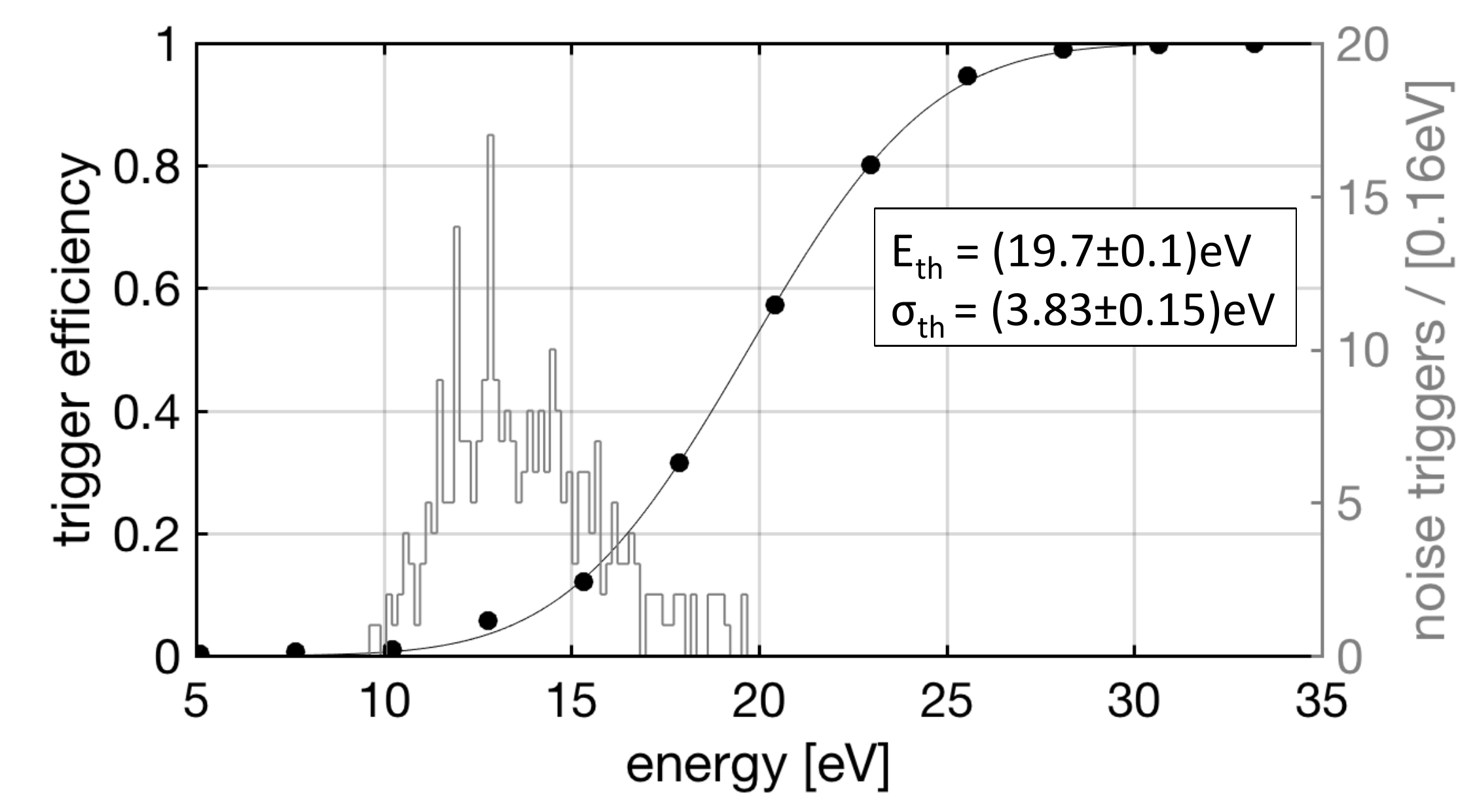}
\caption{Trigger efficiency evaluated at discrete energies (dots, left axis) by using the optimum trigger algorithm. These data points are fitted by an error function (solid line, see text). The histogram shows the distribution of the filter output (maximum) of randomly sampled noise traces (right axis). The  energy threshold $E_{th}$  is chosen such that a negligible amount of noise triggers is accepted. Data  from \cite{cnns_letter}.  }
\label{fig:triggerthreshold}
\end{figure}

The trigger threshold is derived by a systematic study \cite{cnns_letter}. The optimum filter is applied to a set of noise traces, which are randomly recorded by the (standard) DAQ, and is evaluated at every position within a single trace. The maximum filter output for each trace is stored. Fig. \ref{fig:triggerthreshold} shows a histogram of the maximum filter output for $\sim$400 baseline traces. The threshold of the software trigger has to be chosen such that noise triggers are negligible. In the case of this measurement, the trigger threshold is set to 13.0\,mV \cite{cnns_letter}.  To derive the trigger efficiency, artificial template pulses of discrete energies are added to the randomly chosen noise traces. After applying the optimum filter, the fraction of events above and below the trigger threshold are calculated for each simulated energy. The result is shown as dots in Fig. \ref{fig:triggerthreshold}. These data points are well fit by an error function $p_{trig}(E)=0.5\cdot(1+\mathrm{erf}[(E-E_{th})/(\sqrt{2}\sigma_{th})]$ which gives an energy threshold of $E_{th}=(19.7\pm0.1(stat.))$\,eV. This value corresponds to about $5.3\,\sigma_b$ of the baseline noise. As expected, the width of the error function $\sigma_{th}=(3.83\pm0.15)$\,eV is in agreement with the baseline noise $\sigma_b$ evaluated at $E=0$.

To fully exploit this improvement, a software-trigger algorithm for the continuously acquired data was developed.  The trigger threshold was set accordingly to 19.7\,eV.  
The entire data stream is processed with the optimum filter by an algorithm which avoids distortions in the finite Fourier transform (for details see \cite{Piperno:2011fp}). An event is triggered whenever the filter output exceeds the chosen software threshold. Using the same filter for  amplitude evaluation and  triggering guarantees the consistency of the trigger definition and  the energy calibration.
The pulse data from hardware (standard DAQ) and software (continuous DAQ) trigger agree  down to the hardware threshold of $\sim40$\,eV. The pulse height evaluation of the optimum filter and the truncated standard event was studied on an event-by-event basis. In linear region of the pulse response, a maximum deviation of 2.8\,\% is observed  \cite{cnns_letter} which adds to the systematic error of the energy calibration. The final value of the  threshold is $E_{th}=(19.7\pm0.9)$\,eV including statistical and systematic errors. For the following DM analysis we use exclusively the data recorded with the continuous DAQ. 

\subsection{Stability and data selection}
Since the data was not blinded for the calibration measurement \cite{cnns_letter}, we choose the most conservative approach for this  DM analysis:  we apply no data quality cuts which might introduce an energy-dependent efficiency or might bias the result. 


The only cut applied is the so-called stability cut which rejects periods where the detector is not in its correct operating point. The cut is based on the pulse height of control pulses (saturated pulses) which are continuously injected into the ohmic heater every 10\,s. This pulse height should remain constant if the operating point is stable. To be conservative all events within $\pm5$\,min around every outlying control pulse are not used for the analysis. This removes 2.05\,h from the total measuring time of 5.31\,h.

\begin{figure}
\centering
\includegraphics[width=0.5\textwidth]{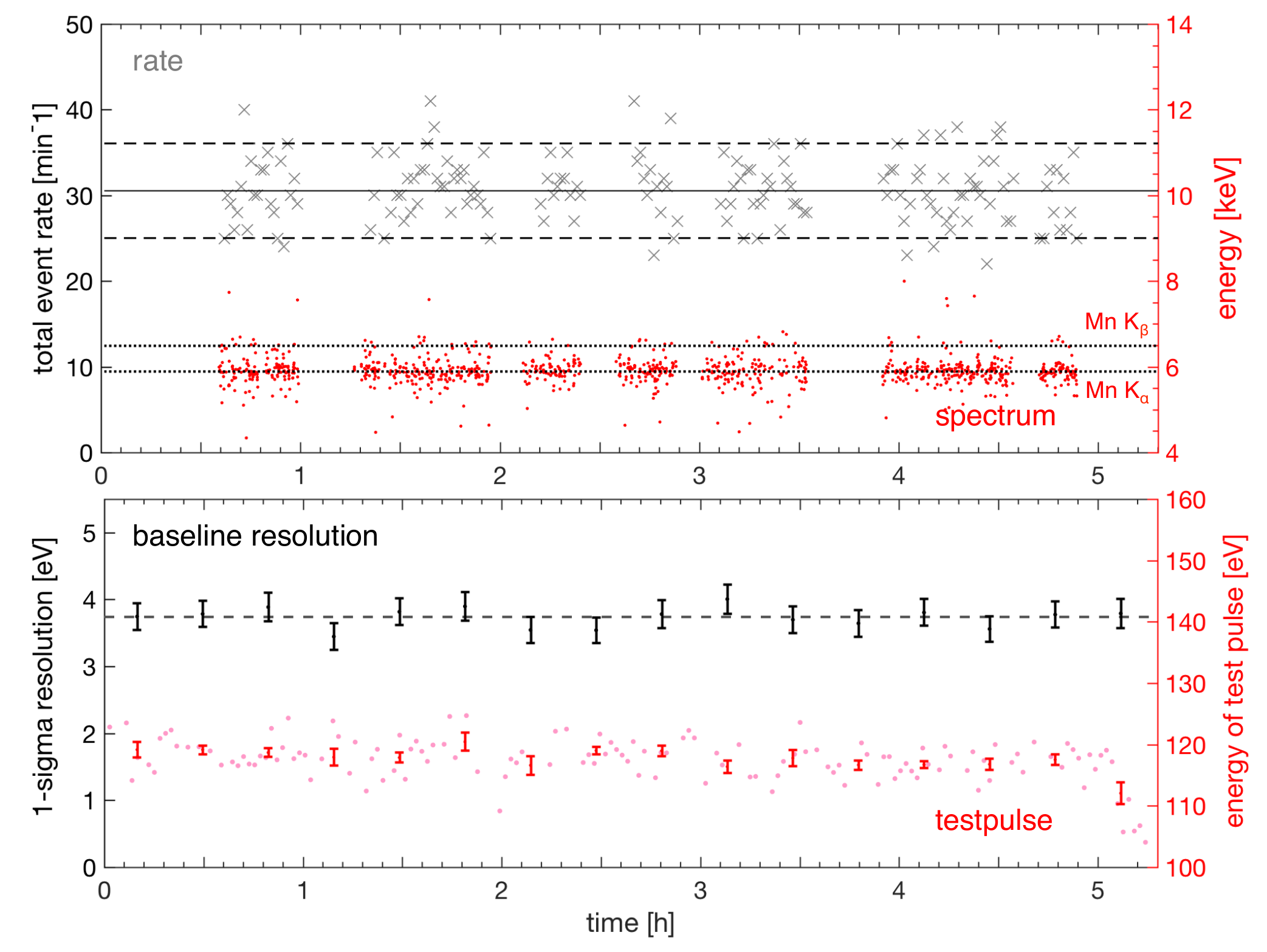}
\caption{Stability of the prototype detector over the measuring time of 5.31\,h after the stability cut. Left y-axis: Total number of counts acquired per minute. Within statistics (dashed lines give Poissonian variance) the rate is stable over the entire run at a mean of 0.51\,Hz (solid line). Right y-axis: Red dots show all events observed in the energy region of 4-8\,keV. The positions of the K$_\alpha$ and K$_\beta$ x-ray energies (literature values) are indicated by dotted lines which are in agreement with the data. This demonstrates the stability of the detector response over time. }
\label{fig:stability}
\end{figure}


Fig.  \ref{fig:stability} shows the stability of various detector parameters over the duration of the measurement. In the top frame the total event rate after the stability cut (grey crosses, left axis) and the energy of all events between 4 and 8\,keV (red dots, right axis) are depicted. The observed event rate is statistically consistent with a constant rate of 0.51\,Hz (solid line). The dashed lines indicate the Poissonian variance (1\,$\sigma$). The dotted lines correspond to the literature values of the K$_\alpha$ and K$_\beta$ lines of $^{55}$Mn. Within statistics the positions of the x-ray lines are stable with time, demonstrating the stability of the energy calibration. The gaps in the data correspond to periods removed by the stability cut.
The bottom frame shows the energy resolution as obtained from empty baseline samples (black error bars, left axis) and the energy of periodically injected test pulses corresponding to an energy of $\sim 120$\,eV (red dots, right axis) together with their average (1$\sigma$ error bars). Over the entire measuring time the variance of the baseline is in agreement with $\sigma_b=(3.74\pm0.21)$\,eV. The test pulse response also remains constant over time demonstrating again the stability of the detector response, except for the period at 5.0\,h when the base temperature of cryostat has changed. This period is however removed by the stability cut (compare to upper frame of Fig. \ref{fig:stability}  ).



\section{Results and discussion}\label{sec:results}
We define the linear region of the pulse response which extends from the energy threshold of 19.7\,eV to 600\,eV as our region-of-interest for DM search (ROI).  The final spectrum is depicted in Fig. \ref{fig:spectrum} (inset) with a binning of 5\,eV. 
All 511 events in the ROI  are considered conservatively as candidate events for DM particle-nucleus scattering, although it is assumed that these pulses originate from backgrounds. Since we do not apply data-quality cuts,\footnote{This implies that the cut efficiency is 1 all over the ROI.}  the remaining events (higher energetic particle pulses, test pulses, control pulses and possible artifacts) are unspecified in the analysis. Again conservatively, we remove the sampling time (614.28\,ms per pulse) of the 5779 events which lie outside the ROI from the live-time. This results in a net live-time of 2.27\,h which corresponds to a net exposure of  0.046\,g-days.

The constant background  level of  $\sim1.2\cdot10^{5}$\,counts/[kg keV day]  is not unexpected due to lack of shielding against ambient radiation and due to the operation of the detector above ground. In addition, an x-ray calibration source was present during the entire measurement. The steep rise of the event rate towards threshold can be caused by source-related Auger electrons. Also, backgrounds such as beta/gamma or alpha decays on surfaces surrounding the detector could  contribute to the exponentially increasing spectrum (see \cite{cnns_letter} for a more detailed discussion). The presence of exponentially rising, unknown backgrounds prohibits an unambiguous discovery of DM, however promising techniques are on their way to significantly reduce and understand remaining backgrounds \cite{cnns_long}.

Since the threshold of the prototype detector is lowered by more than one order of magnitude compared to previous macroscopic devices, the mass range of DM particles below 500\,MeV/$c^2$ can be probed for the first time despite a relatively high background level. An upper limit on the elastic spin-independent DM particle-nucleon cross section is derived. For the  limit calculation we use the Yellin optimal interval method \cite{PhysRevD.66.032005} which was developed to derive DM limits in the presence of backgrounds with unknown energy spectra. The sensitivity is based on a comparison of the observed spectrum (see Fig. \ref{fig:spectrum} (inset)) and the expected recoil spectrum from DM particles of a certain mass $m_\chi$ in Al$_2$O$_3$. 

For the calculation of the recoil spectrum we use the standard astrophysical parameters for the DM halo of the Milky Way: a Maxwellian velocity distribution, an asymptotic velocity of 220\,km/s and a galactic  escape velocity of 544\,km/s. The local DM density at Earth position is assumed to be 0.3\,GeV/cm$^3$. The Helm form factor \cite{Lewin:1995rx} is used in the calculation of the scattering cross section. Both elements of Al$_2$O$_3$ are considered as targets. To take into account the finite energy resolution of the calorimeter the calculated spectrum  is convolved with the Gaussian resolution ($\sigma_b=(3.74\pm0.21)$\,eV at threshold).

\begin{figure}
\centering
\includegraphics[width=0.5\textwidth]{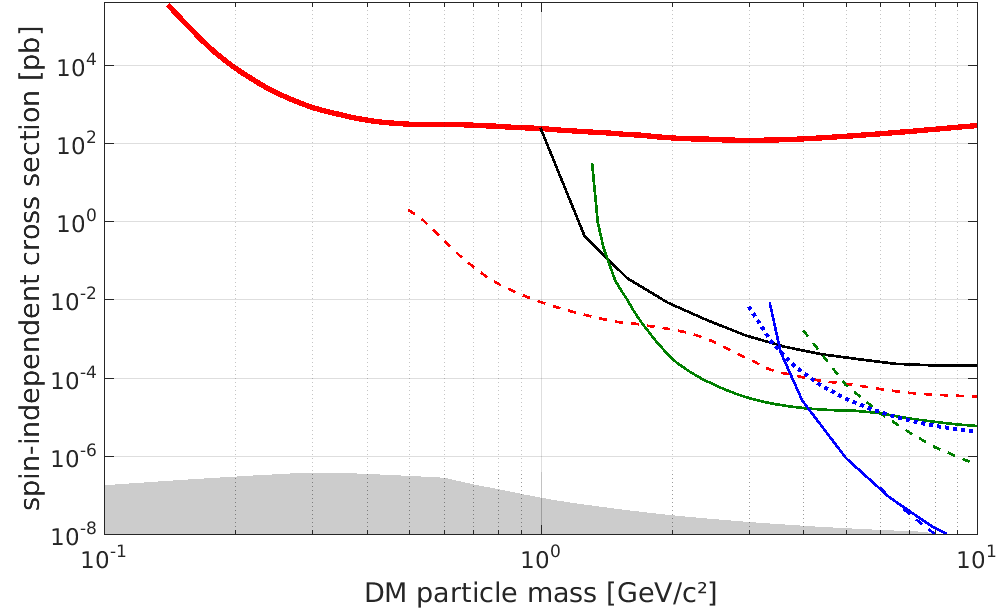}
\caption{Parameter space of DM mass vs. spin-independent DM particle-nucleon  cross-section. The result of this work (solid red) explores a new mass region between 140 and 500\,MeV/c$^2$ and extends the reach of  DM direct searches. In comparison selected experimental results are shown: CRESST-II Phase 2 (dashed red) \cite{Angloher:2015ewa}, CDMS-lite (full green) \cite{Agnese:2015nto}, EDELWEISS (dashed green) \cite{Hehn2016}, XENON100 low mass analyis (dotted blue) \cite{PhysRevD.94.092001}, PandaX (full blue) \cite{PhysRevLett.118.071301}, LUX (dashed blue) \cite{PhysRevLett.118.021303}  and DAMIC (black) \cite{PhysRevD.94.082006}. The shaded grey area indicates the neutrino-floor calculated for Al$_2$O$_3$. }
\label{fig:limit}
\end{figure}

The measurement presented here extends the reach of direct DM search experiments for \textit{l}DM masses to below 500\,MeV/c$^2$. The achieved threshold of $E_{th}=(19.7\pm0.9)$\,eV allows to probe elastic scattering of DM particles down to masses of 140\,MeV/c$^2$. 
 Fig. \ref{fig:limit} shows the upper limit with 90\,\% confidence level achieved in this work (red line). Below 500\,MeV/c$^2$ we explore a new region of parameter space via the direct detection of nuclear recoils and  improve existing limits from the detection of bremstrahlung emission which accompanies  nuclear recoils  \cite{PhysRevLett.118.031803,McCabe:2017rln}. For masses below $\sim1$\,GeV/c$^2$ O-recoils dominate the recoil spectrum, while for higher masses Al-recoils provide the highest sensitivity. The transition is visible as a weak inflection in the exclusion curve. 

In comparison, selected results of direct DM search experiments on the elastic spin-independent DM particle-nucleon cross-section are shown. 

\section{Outlook}
Gram-scale cryogenic calorimeters have demonstrated their high potential for DM search.  An  Al$_2$O$_3$ prototype has explored a new range of DM particles below 500\,MeV/c$^2$. This first result can be improved in two directions: 
1) Detector performance. A recently developed scaling law \cite{cnns_letter} predicts energy thresholds in the 1-10\,eV regime for cubic detectors of $\lesssim1$\,g.  Modifications of the TES sensor are foreseen in order to reach this goal \cite{cnns_letter}. 2) Background level. The shielding of the setup used for the measurement can be improved significantly. We expect a reduction of the background rate by two orders of magnitude when installing a low-radioactivity Pb shielding around the experimental volume.  A 4$\pi$ active cryogenic veto around the calorimeter would significantly reduce surface-related backgrounds (see MC results in \cite{cnns_long}) which are suspected to cause backgrounds at lowest energies. 

The next generation of gram-scale calorimeters which will be developed for the \textit{$\nu$-cleus} experiment \cite{cnns_long} may extend the reach of  direct search experiments to  DM particle masses of $\mathcal{O}$(10\,MeV/c$^2$) and probe new models for DM. Due to hints that backgrounds at lowest energies are not dominated by cosmogenic radiation,  large improvements in sensitivity may be achieved in above-ground or shallow laboratories.

\bibliography{MeVpaper}

\newpage

\end{document}